\documentclass[longauth]{aa} 
\usepackage{graphicx,natbib,txfonts}
\begin{document}
\title{The first orbital solution for the massive  colliding-wind binary HD~93162 ($\equiv$WR~25)
\thanks{Based on observations collected at the European Southern Observatory
(La Silla, Chile), at the Las Campanas Observatory, at the Cerro-Tololo
Observatory (Chile) and at the CASLEO (El Leoncito, Argentina).}}

\author{R. Gamen
\inst{1}\fnmsep\thanks {Visiting Astronomer, CASLEO, 
Argentina, and LCO, Chile}
\and
E. Gosset
\inst{2}\fnmsep\thanks{Research Associate FNRS (Belgium)}
\and
N. Morrell
\inst{3}\fnmsep\thanks {Visiting Astronomer, CASLEO, 
Argentina, CTIO and LCO, Chile}
\and
V. Niemela
\inst{4}\fnmsep\thanks {Member of Carrera del Investigador, 
CIC--BA, Argentina. 
Visiting Astronomer, CASLEO, Argentina, and CTIO, Chile.}  
\and
H. Sana
\inst{2}\fnmsep\inst{5}                                                                        \and
Y. Naz\'e\inst{2}\fnmsep\thanks{Postdoctoral Researcher FNRS (Belgium)}
\and
G. Rauw\inst{2}\fnmsep$^{\star\star\star}$
\and
R.~Barb\'a
\inst{1}\fnmsep\thanks {Member of Carrera del Investigador, 
CONICET, Argentina. 
Visiting Astronomer, CASLEO, Argentina, and LCO, Chile}
\and
G. Solivella
\inst{4}\fnmsep\thanks {Visiting Astronomer, CASLEO, Argentina}
}

\offprints{R. Gamen, \email{rgamen@dfuls.cl}}

\institute{Departamento de F\'{\i}sica, Universidad de La Serena,
           Benavente 980, La Serena, Chile\\
           \email{rgamen@dfuls.cl}
\and
	Institut d'Astrophysique et de G\'eophysique, Universit\'e de Li\`ege,
	All\'ee du 6 Ao\^ut, 17, B-4000, Li\`ege  (Sart Tilman), Belgium\\
          \email{gosset@astro.ulg.ac.be}
\and
           Las Campanas Observatory, The Carnegie Observatories,
	   Casilla 601, La Serena, Chile\\  
	   \email{nmorrell@lco.cl} 
\and  
           Facultad de Ciencias Astron\'omicas y Geof\'{\i}sicas, 
	   Universidad de La Plata, Paseo del Bosque S/N, 1900,   
	   La Plata, Argentina\\  
	   \email{virpi@fcaglp.unlp.edu.ar}  
\and
           European Southern Observatory, Casilla 19001, Santiago 19,
           Chile \\
           \email{hsana@eso.org} 
}  
  
\date{Received $<$date$>$ / accepted $<$date$>$}  
  
\abstract  
{Since the discovery, with the EINSTEIN satellite,
of strong X-ray emission associated with HD~93162 ($\equiv$WR~25),
this object has been predicted
to be a colliding-wind binary system. However, radial-velocity variations  
that would prove the suspected binary nature have yet to be found.   
}  
{We spectroscopically monitored this object to investigate its possible variability 
to address this discordance. 
}  
{We compiled the largest available radial-velocity  
data set for this star to look for variations that might be due to  
binary motion. We derived radial velocities from spectroscopic  
data acquired mainly between 1994 and 2006,  
and searched these radial velocities for  
periodicities using different numerical methods.  
}  
{For the first time, periodic radial-velocity variations
are detected. Our analysis definitively shows that the Wolf-Rayet star WR~25 is an   
eccentric binary system with a probable period of about 208 days.  
}  
{}  
  
  
\titlerunning{The first orbital solution for WR~25}  
\maketitle  
%
  
\section{Introduction}  
  
Massive stars of spectral type O and Wolf-Rayet (WR)
are important objects that play a crucial role  
in the dynamic and chemical evolution of galaxies. 
They are the major source  
of ionizing and UV radiation and, through their huge mass-loss rates,  
they have a strong mechanical impact on their surroundings.  
Despite their importance, our knowledge of these objects and of their
evolution is still fragmentary. The parameters that predominantly  
determine the evolution of a massive star are its mass (as for any star  
in the HR diagram) and its mass-loss rate, although rotation could also
have an important impact \citep{mem05}.  
In this context, massive O+O and WR+O binaries are key objects because  
their binary nature allows us to determine minimum masses from the  
radial velocity orbital solution. 

It is now well established that strong winds of massive stars  
in binary systems collide \citep{1992ApJ...386..265S}
generating hard X-ray emission. For O stars, this comes in addition
to a softer component that is intrinsic to individual objects
\citep{brg97, san06}. The existence of intrinsic emission for
WR stars is still under investigation \citep{osk03, gos05}.
As for colliding-wind binaries, the observed characteristics of the
collision region may contribute to 
constrain the mass-loss rates and, eventually, 
the inclination of those systems (by studying 
the X-ray emission modulation during the orbital cycle).  
Although colliding-wind binaries are interesting objects, 
only a handful of them have been studied in detail, which is  
particularly true for WR+O systems.  
  
The star \object{HD~93162} \citep[WR~25 in][]{huc01}   
is a bright ($V$~=~8.1) galactic Wolf-Rayet 
located in the Carina Nebula region. 
Its binary nature has been a matter of debate  
for many years. 
It has been classified as WN7 + O7 by \citet{smi68} because 
the spectrum appears as a superposition of WN-type emission 
lines and of absorption lines corresponding to 
an O-type star.
The morphology of the blue optical spectrum of WR~25 has been further
discussed by \citet{wal74}. He classified this star as WN6-A 
\citep[see also][]{wal00}
and considered that there was not sufficient 
information to confirm its binary nature, 
since absorption lines intrinsic to a WN star had been observed in the neighbouring 
binary \object{HD~92740}$\equiv$WR~22 \citep{nie73}.
The massive binary status of WR~22 with
absorption and emission lines moving in phase has been confirmed by several
authors \citep{nie73,1979ApJ...228..206C,rau96,1999A&A...347..127S}. 
In later classification systems of 
WR stars \citep[e.g.][]{huc81,smi96}, the presence of absorption
lines has been noted in different ways, e.g.\ ''+abs'' or adding an ''a'', 
indicating that
absorption lines of unknown origin are observed in the otherwise 
emission-line spectrum.

Radial velocity studies of WR-type spectra with 
absorption lines are thus needed to shed light
on the origin of these lines. In the case of WR~25, previously reported
radial velocity studies have not revealed any orbital motion.
From the study of 15 photographic spectra obtained  
during consecutive nights, \citet{1978A&A....68...41M}  
concluded that WR~25 is probably  
a single star. \citet{1979ApJ...228..206C} reached  
a similar conclusion although  
they noticed a larger scatter in the radial velocities  
of WR~25 than that observed   
for other stars in their study (such as 
\object{HD 93131}$\equiv$WR~24 and WR~22 besides its orbital motion). 

Using EINSTEIN observations, \citet{1982ApJ...256..530S} 
found that WR~25 has an abnormally  
high X-ray flux and, later, \citet{1991IAUS..143..105P}  
suggested that this very high X-ray  
luminosity might be caused by colliding winds in a binary system.  
More recently, \citet{raa03} analyzed {\it{XMM-Newton}}
observations and provided further evidence that the X-ray emission
could be indicative of a colliding-wind binary (CWB) 
although its apparent stability has 
been, on the contrary, taken as an argument
against binarity. \citet{pol06} reported significant variability
as detected with {\it{XMM-Newton}} and discussed the 
possibility of periodic variations in the X-ray flux. 
 
\citet{1992ApJ...386..288D} found polarimetric variability  
in WR~25, as well as a   
wavelength dependence of its polarization angle. 
One of the two proposed explanations    
was that this variability may arise in a long-period  
binary system. The alternative explanation
attributed particular properties
to the intervening interstellar medium. 
In his recent catalog, \citet{huc01} classified WR~25 as WN6h+O4f  
considering that the WR emission lines showed evidence of being  
diluted, although no radial-velocity variations had ever been detected.  
  
To investigate the binary status of WR~25,
the massive star research groups of Li\`ege and La Plata  
independently collected high-resolution spectra of this star over the past 10 years. 
The acquisition of high-resolution spectra
of WR stars is rather unusual due to the difficulties presented by
the comparatively broad emission lines typical of these stars.
However, the radial velocities (RVs) measured for each data set showed variations larger  
than the expected errors. No single periodicity could be found
until the data obtained by the two groups were combined.  
In the following, we show how the analysis of the combined spectroscopic   
data set enabled us to reveal the binary nature of WR~25  
and to derive a preliminary orbital solution for this system.   
  
  
\section{Observations and data reduction}  
  
\begin{table}  
\caption{Observing runs. The first column gives the month and year of the
observations, the second column lists the instrumental
combination utilized. The third column gives the number of spectra.}  
\begin{tabular}{c c c }  
\hline\hline\noalign{\smallskip}  
Date  & Instr. configuration  & n \\  
\noalign{\smallskip}\hline\noalign{\smallskip}  
\multicolumn{3}{c}{High-resolution spectra}\\  
\noalign{\smallskip}\hline\noalign{\smallskip}  
Feb-97& Echelle-REOSC, 2.15-m, CASLEO& 2 \\ 
Mar-97& CES+LC, CAT 1.4-m, ESO       & 4 \\ 
Feb-98& Echelle-REOSC, 2.15-m, CASLEO& 5 \\ 
Feb-99& Echelle-REOSC, 2.15-m, CASLEO& 3 \\ 
May-01& FEROS, 1.5-m, ESO            & 4 \\ 
Mar-02& FEROS, 1.5-m, ESO            & 4 \\ 
Mar-02& EMMI-REMD/Ech, NTT, ESO      & 2 \\ 
Apr-02& FEROS, 1.5-m, ESO            & 1 \\ 
Jan-03& Echelle-REOSC, 2.15-m, CASLEO& 4 \\ 
Mar-03& Echelle-REOSC, 2.15-m, CASLEO& 3 \\ 
Apr-03& Echelle-REOSC, 2.15-m, CASLEO& 5 \\ 
May-03& FEROS, 2.2-m, ESO            & 3 \\ 
Dec-03& Echelle-REOSC, 2.15-m, CASLEO& 2 \\ 
May-04& FEROS, 2.2-m, ESO            & 3 \\ 
Apr-05& Echelle, 2.5-m, LCO   & 4 \\ 
Feb-06& Echelle, 2.5-m, LCO   & 1 \\ 
\noalign{\smallskip}\hline\noalign{\smallskip}  
\multicolumn{3}{c}{Medium-resolution spectra}\\  
\noalign{\smallskip}\hline\noalign{\smallskip}  
Feb-94& 2D-Frutti, 1-m, CTIO         & 2 \\ 
Apr-96& B\&C, 1.5-m, ESO             & 2 \\ 
Feb-02& CSpec, 1.5-m, CTIO           & 4 \\ 
\noalign{\smallskip}\hline\noalign{\smallskip}  
\multicolumn{3}{c}{Low-resolution spectra}\\  
\noalign{\smallskip}\hline\noalign{\smallskip}  
May-73& Cass, 0.9-m, CTIO            & 2 \\ 
Dec-97& B\&C, 2.15-m, CASLEO         & 4 \\ 
Apr-01& REOSC-DS, 2.15-m, CASLEO     & 6 \\ 
\noalign{\smallskip}\hline  
\end{tabular}  
\label{runs}
\end{table}  
  
Our observational data set consists of 50 high-, 8 medium- and 12 
low-resolution spectra   
obtained, from 1994 to 2006 (except for two spectra acquired in 1973),  
at different observatories and with various instrumental configurations.

Comparison (wavelength calibration)
spectra were obtained at the same telescope positions
as the stellar spectra immediately after or before the science exposures,
except for FEROS which is stable enough that three to four
calibrations over the night are sufficient.

The summary of the observations is presented in Table~\ref{runs} where
we give the dates of the run,  the
instrumentation used, and the number of spectra obtained.

\subsection{High-resolution spectra (HRS)}  
 
Twenty four echelle spectra were obtained with the REOSC Cassegrain 
spectrograph 
(on long-term loan from the University of Li\`ege)
attached to the 2.15-m reflector at the Complejo  
Astronomico El Leoncito (CASLEO\footnote{CASLEO is operated under agreement  
between CONICET, SECYT, and the National Universities of La Plata,  
C\'ordoba and San Juan, Argentina.}), using a Tek $1024 \times 1024$ pixel   
CCD as the detector. These spectra cover an approximate wavelength range   
from 3600 to 6000 \AA,  with a spectral resolving power of 26000.
Typical S/N ratios in the continuum 
range between 65 and 100.

Four spectra were acquired with 
the ESO 1.4-m Coud\'e Auxiliary Telescope
feeding the CES spectrograph equipped with the Long Camera and
CCD ESO\#38, yielding an effective resolving power of 65000, and
a wavelength domain between 4035 \AA\ and 4080 \AA .
Only the blue path was used and the exposure time
was between 30 min and one hour, implying spectra with S/N ratios of 100. 
Normalization to the continuum
was performed through the use of a metal-poor star observed under 
similar conditions to WR~25. 

Fifteen spectra covering the whole optical range
(3750-9000 \AA ) were obtained with the Fiber-fed
Extended Range Optical Spectrograph (FEROS), an echelle spectrograph
mounted at the ESO 1.5-m telescope at La Silla and then transferred
to the ESO/MPI 2.2-m telescope in October 2002. 
The detector was a 2k$\, \times \,$4k EEV CCD
with a pixel size of 15~$\mu$m$\, \times \,$15~$\mu$m. The spectral resolving
power of FEROS is 48000. Typical exposure times were between
10 and 20 min according to weather conditions, resulting in typical
S/N ratios in the continuum
between 150 and 200 at the 1.5-m and around 250 at the 2.2-m.

Two spectra were acquired with EMMI attached to the Nasmyth
focus of the NTT telescope. The instrument was used in the echelle
spectrographic mode (REMD-echelle) with grating \#9 and grism \#3,
providing a resolving power of 7700 and 18 usable spectral orders,
covering the wavelength domain from 4040 \AA\ to 7670 \AA . Typical
S/N ratios are around 120.
  
Five spectra were observed with the 2.5-m du Pont telescope at Las Campanas   
Observatory (LCO), Chile, using the echelle spectrograph which  
provides simultaneous wavelength coverage from $\sim$ 3700 to 7000 \AA\  
at a typical resolving power of $\sim$40000.   
The detector is a Tek5 2k $\times$ 2k CCD with pixel size of 24 $\mu$m.
We used a $1 \times 4$ arcsec slit. These spectra have S/N ratios around 100.

\subsection{Medium-resolution spectra (MRS)}

Two medium-resolution spectra (S/N $\sim$ 100)
were obtained in 1996 with the ESO 1.5-m telescope
equipped with the modified Boller \& Chivens spectrograph at La Silla.
The configuration utilized is described in \citet{gos01}.
Four medium-resolution spectra were obtained with the Cassegrain
Spectrograph at the CTIO\footnote{CTIO is operated by the AURA Inc.
under a cooperative agreement with the National Science Foundation as part of 
the NOAO.} 1.5-m telescope in February 2002. A 600
l/mm grating blazed at 3375 \AA\ in second order, combined with a
Loral 1k CCD provided a 2 pixel resolution of 1.5 \AA\ (wavelength
range 3270-4180 \AA ).
Two spectra were observed with the Shectman/Heathcote two-dimensional,
photon-counting detector (2D-frutti) on the Cassegrain spectrograph  at the
CTIO 1-m telescope. The wavelength coverage was from 3800 to 5000 \AA\ 
at a 3 pixel resolution of 1.5 \AA.
Typical S/N ratios range between 50 (2D-frutti data) 
and 250 (CSpec at CTIO). Typical resolving powers for these
MRS are in the range 2000-4000.

\subsection{Low-resolution spectra (LRS)} 

We have also obtained lower resolution spectra with resolving powers 
$\sim$ 1000.

Six LRS were acquired at CASLEO with the
above-mentioned REOSC spectrograph but in its simple dispersion mode.   
This configuration   
provides a sampling of 1.64 \AA\ per pixel, on the Tek 1024$\times$1024 CCD.   

We also used 4 spectra observed with the Boller \& Chivens
spectrograph attached to the 2.15-m telescope at CASLEO. 
A PM~512$\times$512 pixel CCD, with pixel size of 20~$\mu$m, was used
as the detector. The reciprocal dispersion was $\sim2.3$\,\AA\,pixel$^{-1}$,
and the wavelength region observed was about $\lambda\lambda$ 3800 -- 4800~\AA.
Typical S/N ratios of CASLEO LRS are about 200.

In addition, we used two previously unpublished radial velocity values
determined from spectra observed in May 1973 at CTIO with the
Cassegrain spectrograph attached to the 0.9-m telescope. These
spectra were recorded on photographic plates, have a reciprocal
dispersion of 120 \AA /mm, and were widened to 1 mm for a better
visibility of the spectral lines. The spectrograms were measured
for the determination of RVs with a Grant oscilloscope
engine. Because the radial velocity values 
of the N\,{\sc iv} $\lambda$4058 emission 
determined from these two spectra were rather more negative than the mean
of the other 52 observations, they were deemed suspect and were not
included in the radial velocity study published by \citet{1979ApJ...228..206C}.
We note, however, that the radial velocity of the interstellar
Ca\,{\sc ii} absorptions in the two 
photographic spectra ($-32$ km~s$^{-1}$) does
not deviate much from the values for these lines in our digital
spectra (see below).
  
\section{The radial velocity analysis}  

\begin{table}
\caption{Radial velocities (RVs) of the
N\,{\sc iv} $\lambda$4058 emission line measured in the spectra of WR~25. 
The velocities are expressed in the heliocentric rest frame;
time is given in Heliocentric Julian Date.}
\setlength{\tabcolsep}{1.8mm}
\begin{tabular}{rrcrrc}
\hline \hline
HJD & RV   & Dataset & HJD & RV   & Dataset\\
2,400,000+  & km~s$^{-1}$    &&2,400,000+  & km~s$^{-1}$\\
\hline
41823.657 & -112 & LRS & 52328.707 & -18 & MRS \\
41828.620 & -117 & LRS & 52328.718 & -20 & MRS \\
49406.668 & -9 & MRS & 52333.848 & -20 & MRS \\
49409.767 & -23 & MRS & 52335.652 & -18 & HRS \\
50181.671 & -32 & MRS & 52337.685 & -27 & HRS \\
50182.637 & -25 & MRS & 52338.636 & -20 & HRS \\
50505.846 & -47 & HRS & 52339.639 & -22 & HRS \\
50507.802 & -48 & HRS & 52353.616 & -20 & HRS \\
50531.535 & -68 & HRS & 52353.620 & -20 & HRS \\
50531.584 & -70 & HRS & 52383.507 & -48 & HRS \\
50532.513 & -59 & HRS & 52655.865 & -29 & HRS \\
50534.524 & -69 & HRS & 52657.870 & -29 & HRS \\
50807.849 & -15 & LRS & 52658.861 & -30 & HRS \\
50809.868 & -17 & LRS & 52659.855 & -29 & HRS \\
50811.869 & -28 & LRS & 52710.711 & -4 & HRS \\
50812.856 & -26 & LRS & 52711.827 & -6 & HRS \\
50842.800 & -7 & HRS & 52712.832 & -9 & HRS \\
50846.791 & -7 & HRS & 52735.672 & -11 & HRS \\
50847.877 & -7 & HRS & 52736.673 & -18 & HRS \\
50850.855 & -11 & HRS & 52737.631 & -15 & HRS \\
50852.798 & -8 & HRS & 52738.590 & -12 & HRS \\
51209.865 & -15 & HRS & 52739.658 & -22 & HRS \\
51216.881 & -5 & HRS & 52782.493 & -32 & HRS \\
51218.880 & -10 & HRS & 52783.498 & -30 & HRS \\
52005.575 & -80 & LRS & 52784.493 & -30 & HRS \\
52007.577 & -102 & LRS & 52985.841 & -42 & HRS \\
52008.576 & -110 & LRS & 52989.842 & -37 & HRS \\
52009.612 & -81 & LRS & 53131.510 & -11 & HRS \\
52011.635 & -103 & LRS & 53133.605 & -10 & HRS \\
52013.589 & -74 & LRS & 53135.485 & -11 & HRS \\
52037.638 & -16 & HRS & 53480.710 & -46 & HRS \\
52038.586 & -20 & HRS & 53481.584 & -39 & HRS \\
52039.618 & -22 & HRS & 53488.559 & -28 & HRS \\
52040.622 & -15 & HRS & 53490.662 & -26 & HRS \\
52328.697 & -22 & MRS & 53772.679 & -12 & HRS \\
\hline
\end{tabular}
\label{tableRVs}
\end{table}

We have determined the radial velocities of WR 25 measuring the position of
the N\,{\sc iv} $\lambda$4058 emission
line because this line is narrow and strong enough to minimize
measurement errors. In addition, it is expected to be formed
relatively deep in the WR wind, and thus to better reflect
the true motion of the WR star. We arbitrarily
adopted for that line the rest wavelength 
$\lambda_0$=4057.9~\AA\ as given by
\citet{cll77}.
The adopted method consists of fitting Gaussian profiles to the
observed line using either MIDAS or IRAF routines.
We worked on spectra normalized to the continuum and, to homogenize
the RV measurements, we favoured the position as measured 
on the upper part of the line; this has the advantage of being
less dependent on the errors in the definition of the
continuum. 
The RVs of the N\,{\sc iv} $\lambda$4058 emission line as measured 
in the spectra of WR~25 are reported in Table~\ref{tableRVs}.
These values should be cautiously considered as preliminary because
various effects (e.g.\ intrinsic line-profile variations)
could render the measured RVs inaccurate.
On a few good spectra, we used other techniques to measure 
the position of the N\,{\sc iv} $\lambda$4058 line, e.g.\
one based on the lower part of the emission profile. 
The resulting RVs differ from the adopted RVs
measured on the basis of the upper part
by 0.06 \AA\ or 4.5 km\,s$^{-1}$ at maximum. 
No systematic shift has been detected. These results give
further support to the reported RVs and provide an idea of the error
on the measurement.  

\begin{figure}
   \centering
   \includegraphics[width=8.5cm]{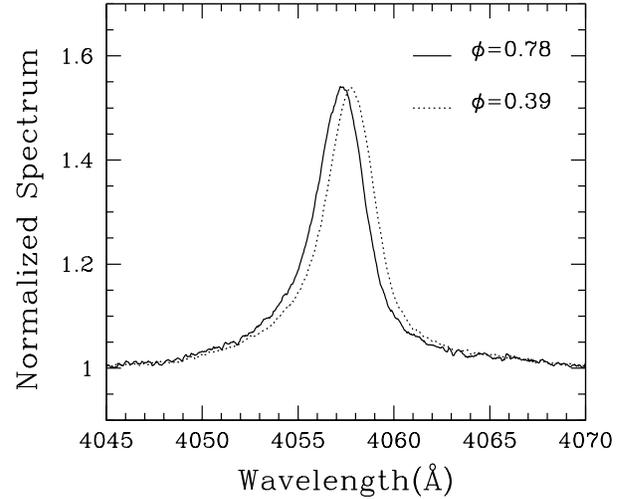}
   \caption{Two typical line profiles corresponding to high
S/N FEROS spectra of WR~25. The emission line is N\,{\sc iv} $\lambda$4058 and
is shown in the heliocentric rest frame.
One (plain line) was observed on HJD\,2\,452\,383.507 (corresponding to $\phi$~=~0.78,
see Table~\ref{orbital}) and is exhibiting a RV of --48 km\,s$^{-1}$; the other was
observed on HJD\,2\,453\,133.605 ($\phi$~=~0.39) corresponding to a velocity
of --10 km\,s$^{-1}$. The main change between the two spectra is a shift in RV.
}
\label{emission}
\end{figure}

In Fig.~\ref{emission}, we show
two typical line profiles. One was observed on
HJD\,2\,452\,383.507 (corresponding to $\phi$~=~0.78, see below) and
exhibits a RV of --48 km\,s$^{-1}$. The other, acquired on
HJD\,2\,453\,133.605 ($\phi$~=~0.39), corresponds to a velocity
of --10 km\,s$^{-1}$. These spectra are representative of two
extreme positions of the line as observed by us at high resolution.
The main difference between the two spectra of Fig.~\ref{emission} is a Doppler shift and
line-profile variations, if they exist, are only second
order effects. On the other hand, the smallness of the shift
compared to the line width explains the difficulty
in performing fully accurate measurements through a fit of the entire
profile.

\begin{figure}  
   \centering  
   \includegraphics[width=8.5cm]{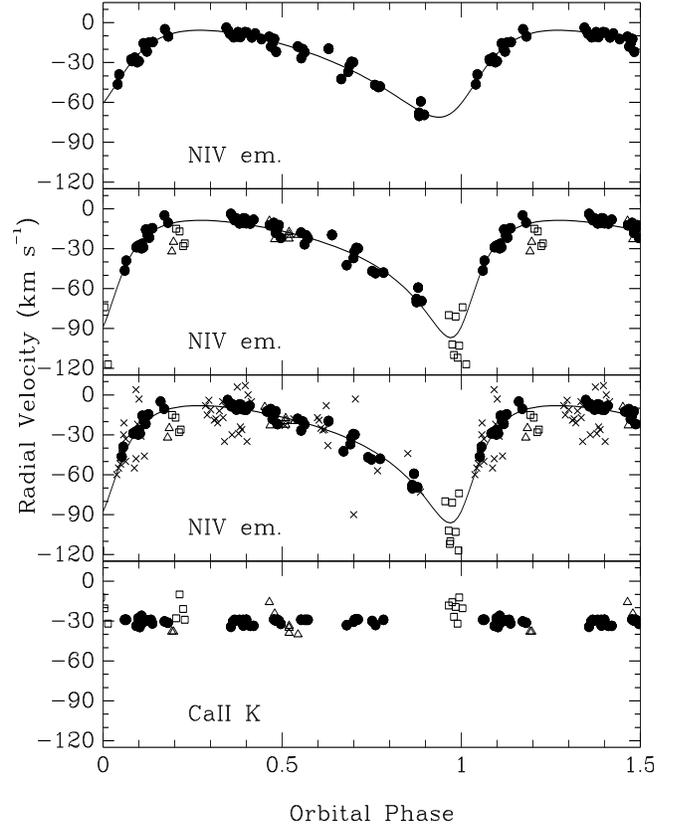}
   \caption{  
Observed radial velocities (along with the fitted models) corresponding
to the 
N\,{\sc iv} $\lambda$4058 emission line in the  
spectra of WR~25. The data are phased according to the ephemeris 
as given in Table~\ref{orbital}; phase 0.0 corresponds to periastron. 
The different panels
illustrate different analyzed data sets.  
{\it{Top}}:   
Our HRS data only (filled circles).  
{\it{Middle top}}: HRS+MRS+LRS data; 
open triangles represent the RVs in the MRS
set, and squares the LRS ones.  
{\it{Middle bottom}}:
All + published data; crosses are the RVs from \citet{1979ApJ...228..206C},  
\citet{1978A&A....68...41M}, and \citet{1995A&AS..113..459H}. 
In each plot, the solid curve represents the relevant orbital solution
as given in Table~\ref{orbital}.  
{\it{Bottom}}: Radial velocities of the interstellar Ca\,{\sc ii} K absorption feature measured on our spectra.
}  
\label{RVs}  
\end{figure}

We also measured the Ca\,{\sc ii} $\lambda$3933 interstellar absorption  
line. In the echelle spectra, this line presents at least 5 distinct  
components. Therefore, we measured the position of the central line of   
the three main components. Averaging the RVs obtained in the echelle   
spectra, we derived a  mean of $-30\pm$4~km~s$^{-1}$. In  
Fig.~\ref{RVs} (bottom), we also show the RVs of this 
Ca\,{\sc ii} line measured 
in our WR~25 spectra. This confirms the good agreement existing
among the different instrumental configurations used in this work.

Having noticed that the position of the N\,{\sc iv} $\lambda$4058 line
was variable, we studied the time series of the measured RVs. 
We searched for periodicities using two independent methods:
the algorithm to derive periods of cyclic phenomena described in \citet{mar80},  and 
a Fourier-type analysis method by 
\citet[][see also comments by Gosset et al., 2001]{hmm85}. 

\begin{figure}  
   \centering  
   \includegraphics[width=6.7cm, angle=-90]{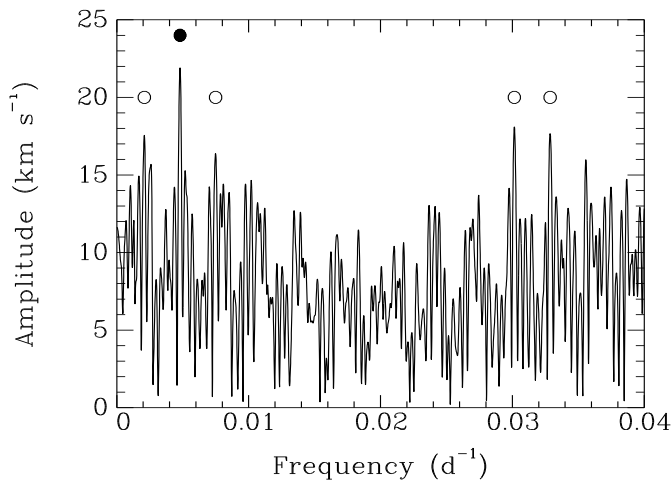}
   \caption{
Amplitude Spectrum (square root of the power spectrum) as a function
of the frequency expressed in d$^{-1}$ and corresponding to the analysis
of the HRS data set. From 0.04~d$^{-1}$ to 0.5~d$^{-1}$, no peak
exceeds 13 km\,s$^{-1}$. The adopted main frequency is marked by
a filled circle and the expected positions of the main aliases of
this frequency are marked with open circles.
}  
\label{fourier}  
\end{figure}

\begin{figure}  
   \centering  
   \includegraphics[width=6.7cm,angle=-90]{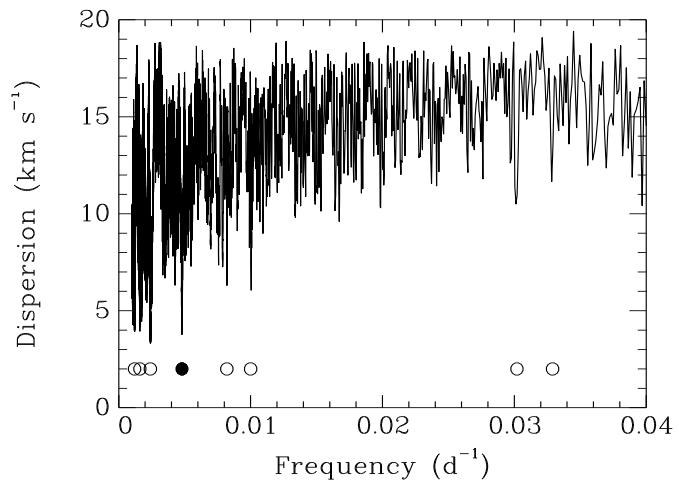}
   \caption{
Run of the \citet{mar80} statistic as a function
of the frequency expressed in d$^{-1}$ and corresponding to the analysis
of the HRS data set. The main dips are seen at
$\nu$~=~0.0012~d$^{-1}$,
$\nu$~=~0.0016~d$^{-1}$,
$\nu$~=~0.0024~d$^{-1}$ and 
$\nu$~=~0.0048~d$^{-1}$; the first three are subharmonics of
$\nu$~=~0.0048~d$^{-1}$ and are a well-known artifact of this kind of
method. Other dips are visible at
$\nu$~=~0.0082~d$^{-1}$,
$\nu$~=~0.0100~d$^{-1}$,
$\nu$~=~0.0302~d$^{-1}$ and $\nu$~=~0.0329~d$^{-1}$.
The last two are also aliases of the $\nu$~=~0.0048~d$^{-1}$ dip. The cited
frequencies are marked as in Figure~\ref{fourier}.
}  
\label{marmuz}  
\end{figure}

We first analyzed the data set consisting of the 50 high-resolution
spectra (HRS). The particular distribution of the observing times
induces some aliasing at $\delta \nu$~=~0.00045~d$^{-1}$, at
$\delta \nu$~=~0.00270~d$^{-1}$, at
$\delta \nu$~=~0.02540~d$^{-1}$ and at
$\delta \nu$~=~0.02810~d$^{-1}$ as revealed by e.g.\ the spectral
window \citep[see the definition in][]{deem75}. 
The Amplitude Spectrum (square root of the
power spectrum) of the HRS data
is given in Fig.~\ref{fourier}. The highest ordinate
is located at $\nu$~=~0.004795~d$^{-1}$. The two neighbouring
peaks are aliases of this frequency; this is also the case
for the peaks at $\nu$~=~0.00209~d$^{-1}$, at $\nu$~=~0.00749~d$^{-1}$,
at $\nu$~=~0.03015~d$^{-1}$ and at $\nu$~=~0.03286~d$^{-1}$.
All of them thus belong
to the same family and we consider that the dominant frequency
is the one at $\nu$~=~0.004795~d$^{-1}$.
This is further confirmed by a
decomposition of the RV curve into several frequencies using the
multifrequency approach \citep[see Eqs. A11 to A19 in][]{gos01}.
The main frequency varies from $\nu$~=~0.004795~d$^{-1}$
to $\nu$~=~0.004805~d$^{-1}$ depending on the secondary
frequencies being inventoried: we thus adopt
$\nu$~=~0.00480~d$^{-1}$ ($P$~=~208.3~d) as the true progenitor.
We estimated the error on the period by taking as a conservative
upper limit one tenth of the peak width (which is a function
of the time baseline); we derived $\sigma_{\nu}$~=~0.00003~d$^{-1}$
and thus $\sigma_{P}$~=~1.3~d. 
The Marraco \& Muzzio method applied to the 
RVs measured on the HRS spectra   
gave as a most probable   
period $P$~=~208.3$\pm$0.5 d along with some aliases, in good agreement
with the Fourier results (see Fig.~\ref{marmuz}).

A phase diagram with the adopted period presents a clear gap
(around phase 1.0, see Fig.\ref{RVs}).
Therefore, we complemented our HRS data with other data sets.
We thus defined two subsequent data sets:
\begin{itemize}   
\item our 50 RVs measured on the high-resolution spectra (HRS)   
complemented by the 8 medium-resolution data (MRS) and by
the 12 low-resolution spectra (LRS), thus 70 data points labelled
HRS+MRS+LRS,   
\item all our 70 RVs plus the 54 published by \citet{1978A&A....68...41M} and 
\citet{1979ApJ...228..206C}, complemented with the RV measured in one 
spectrum provided by \citet{1995A&AS..113..459H}.   
\end{itemize}

We also ran the period-search algorithms using these enlarged data sets.
In the HRS+MRS+LRS data set, the Fourier analysis favoured
a frequency $\nu$~=~0.004820~d$^{-1}$ which is within the error bars
of the previously quoted one. Finally, the data set
all+published indicates a frequency
$\nu$~=~0.004810~d$^{-1}$ ($\sigma_{\nu}$~=~7.7\,10$^{-6}$~d$^{-1}$)
corresponding to $P$~=~207.9~d ($\sigma_{P}$~=~0.3~d). 
The \citet{mar80} method applied to the all+published data set  
similarly gives $P$~=~207.8$\pm$0.3~d.

To further refine the period determination, we ran   
GBART, an improved version\footnote{Available upon request from
\newline  
{\tt{ftp://lilen.fcaglp.unlp.edu.ar
\newline
/pub/fede/gbart-0.1-41.tar.gz}}}  
of the orbital solution 
program originally published by \citet{ber68},
with the three above-mentioned data sets.  
The RVs measured on HRS were weighted   
with 1, MRS with 0.5, and LRS or published, with 0.1.   
The orbital parameters obtained for each data set are 
explicitly given  
in Table~\ref{orbital} and the corresponding orbital
solutions are depicted in Fig.~\ref{RVs}.
We obtained quite eccentric   
orbits with $e$ ranging between 0.35 and 0.5,
orbital semi-amplitudes $K$ between  
33 and 44 km~s$^{-1}$, and 
different values for the orbital period.
%
\begin{table}  
\caption{Orbital solutions corresponding to the RVs of 
the N\,{\sc iv} $\lambda$4058 emission line within different  
data sets. The quoted errors correspond to 1$\sigma$ uncertainties.
Symbols have the canonical meaning. The last three correspond,
respectively, to the mass function, the standard deviation of the fit
and the number of data points involved.}   
\label{orbital}  
\setlength{\tabcolsep}{0.7mm}  
\begin{tabular}{l lrcl lrcl lrcl}  
\hline\hline\noalign{\smallskip}
&& \multicolumn{3}{c}{HRS only} & & \multicolumn{3}{c}{HRS+MRS+LRS} & & \multicolumn{3}{c}{All + published}\\
\noalign{\smallskip}\hline\noalign{\smallskip}
$P$ [d] &  & 208.3 & $\pm$ & 0.2 &  & 207.9 & $\pm$ & 0.1 & & 207.85 & $\pm$ & 0.02\\
$V_{0}$ [km\,s$^{-1}$] &  & --30.6 & $\pm$ & 0.7 &  & --34.2 & $\pm$ & 0.7 & & --34.6 & $\pm$ &0.5\\
$K$ [km\,s$^{-1}$] &  & 33 & $\pm$ & 2 &  & 42 & $\pm$ & 2 & & 44 & $\pm$ &2\\
$e$ &  & 0.35 & $\pm$ & 0.03 &  & 0.48 & $\pm$ & 0.02 & & 0.50 & $\pm$ & 0.02\\
$\omega$ [degrees] &  & 227 & $\pm$ & 4 &  & 216 & $\pm$ & 3 & & 215 & $\pm$ & 3\\  
$T_{\rm Periast}$ [d]$^{\ast}$ &  & 1598 & $\pm$ & 3 &  & 1597 & $\pm$ & 2 && 1598& $\pm$ &1\\
$T_{\rm RV max}$ [d]$^{\ast}$ &  & 1654 & $\pm$ & 3 &  & 1655 & $\pm$ & 2 &&1654& $\pm$ &1\\
$a\,\sin i$ [R$_\odot$] &  & 125 & $\pm$ & 8 &  & 151 & $\pm$ & 9 && 156 & $\pm$ &8\\
$F(\mathcal{M})$ $[$M$_\odot]$ &  & 0.6 & $\pm$ & 0.1 &  & 1.1 & $\pm$ & 0.2 && 1.2& $\pm$ &0.2\\
$\sigma$ [km\,s$^{-1}$] &  & 2.5 &  &  &  & 3.7 & & & & 4.7 &   &  \\
n  &  & 50 & &  &  & 70 & & & & 124 & &  \\
\noalign{\smallskip}\hline\noalign{\smallskip}
\multicolumn{13}{l}{$\ast$: Heliocentric Julian Date 2,450,000+} \\
\end{tabular}
\end{table}
%

The fact that there is no HRS RV more negative than \mbox{--70 km~s$^{-1}$},
while some LRS RVs indicate --110 km~s$^{-1}$,
is intriguing. Clearly the HRS data set presents a strong gap around phase 1.0. 
This gap can be filled by LRS data obtained in May 1973 and April 2001.
Even though the errors on these RVs are expected to be larger and despite
the low weight given, these LRS data have a large impact on the
eccentricity and on the orbital semi-amplitude derived.
Adding the published data to ours still yields an orbital solution
that remains in good agreement with the HRS+MRS+LRS one.
However, none of the published
RVs have been acquired near the critical phase (1.0).

Although the exact orbital solution remains uncertain, our various
data sets allow us to conclude that WR~25 presents RV variations
with a period of about 208 days. These variations
are indicative of an eccentric binary system. Strictly
periodic RV variations with such a long time-scale are difficult
to interpret in other terms.
The exact values for some of the parameters are highly dependent
on low resolution data. Therefore, we need
further good quality high-resolution data in order to derive a fully
definitive orbital solution for the WR primary.

We also considered other lines in the HRS spectra of WR~25. 
Clearly, certain lines such as N\,{\sc v} $\lambda \lambda$4603-4619
vary in agreement with the derived orbital solutions. 
However, some absorption lines such as
He\,{\sc i} $\lambda$5876 and $\lambda$4471 seem to vary in anti-phase,
plausibly revealing the signature of the binary companion.

Studying other lines will help to verify whether
the RVs derived from the motion of N\,{\sc iv} $\lambda$4058
represent the orbital motion
of the primary; this can not be established on the basis of a
single line. This is beyond the scope of the present paper and requires
high-resolution high S/N data covering all the phases. 
  
\section{Conclusions}  
  
We have found that the RVs of the N\,{\sc iv} $\lambda$4058 emission line in the  
spectrum of WR~25 show variations with a period of about $P$=208 days.  
These variations indicate that the WN star WR~25 is an eccentric binary system.  

The He\,{\sc i} $\lambda$5876 and $\lambda$4471 absorption  
lines seem to show an anti-phase motion and should thus
partly belong to the companion which is supposed to be of the OB type.
  
The companion seems to be fainter than the WR primary, indicating that  
WR~25 is perhaps another example of a young binary system where the  
WR component is the most massive star. The prototype of this class  
is WR~22 \citep[WN7h+O9III-V,][]{huc01} which turned out to be a 
massive binary system  (55 M$_\odot$ + 21 M$_\odot$) in an 80-day period   
orbit \citep{rau96, 1999A&A...347..127S}.  
The primary of WR~22 is most probably an example of the  
long-searched for case of a core H burning star resembling  
a WR star due to its luminosity and mass. The star is on its way to  
becoming a hydrogen-free WR-type star. These objects are extremely rare  
and thus interesting to study in detail. 
  
WR~25 is one of the brightest WN stars in the  
X-ray domain, too bright to be explained by intrinsic  
X-ray emission. It thus was customarily classified as a  
putative colliding-wind binary. Fig.~\ref{RVs} is the first proof that  
WR~25 is a binary as suspected from its X-ray emission.  
On the basis of the existing {\it{XMM-Newton}} data,    
\citet{pol06} found   
that the X-ray emission of WR~25 is variable although the period
cannot be derived independently from the X-ray data alone.  
WR~25 was brighter in the X-ray domain on JD\,2\,452\,842.6 which according to our
ephemeris corresponds to a phase of 0.96-0.98.  The star thus seems to be  
brighter at periastron as predicted for colliding-wind binaries in the adiabatic 
regime \citep[see e.g.][]{1992ApJ...386..265S}.
Although this should be further investigated, WR~25  appears to be an example of a 
colliding-wind binary system.   
  
We are currently planning to monitor the optical spectrum of WR~25 around  
the expected time of minimum radial velocity of the WR 
emission lines in order to derive a full definitive orbital solution.
High-resolution high S/N ratio data are essential. 
These spectra will help to improve the orbital solution, and to detect the   
absorption lines of the companion, which for this particular  
phase domain should be
located to the red of the WN emission line, thus avoiding confusion with
the absorption components of  P-Cygni profiles from the primary.
  
\begin{acknowledgements} 
The Li\`ege team is greatly indebted to the Fonds National
de la Recherche Scientifique (FNRS, Belgium) for support.
This research is supported in part by contract P5/36 IAP (Belspo) and
through the PRODEX XMM and INTEGRAL contracts. 
The La Plata team thanks the Directors and staff of CASLEO, LCO, and CTIO for 
the use
of their facilities, and acknowledges the use, at CASLEO, of the CCD and
data acquisition system supported under US NSF grant AST-90-15827 to R.M. Rich.
We also thank Nolan Walborn for his useful comments on this work, and
Mariela Corti for kindly obtaining 3 spectra for us.
RB acknowledges the support of FONDECYT Program No 1050052.
This research has also received partial 
financial support from IALP, CONICET, Argentina.
\end{acknowledgements}

\end{document}